\documentclass{article}

\usepackage{graphicx}

\usepackage{amssymb}
\usepackage{natbib}

\begin{document}



\title{Analytical benchmark for non-equilibrium radiation diffusion in finite size systems}


\author{Karabi Ghosh\footnote{email: karabi@barc.gov.in}\\ \\ Theoretical Physics Division,\\
 Bhabha Atomic Research Centre, Mumbai 400085, India}
\date{}
\maketitle
\begin{abstract}
Non-equilibrium radiation diffusion is an important mechanism of energy transport in Inertial Confinement Fusion, astrophysical plasmas, furnaces and heat exchangers. In this paper, an analytical solution to the non-equilibrium Marshak diffusion problem in a planar slab and spherical shell of finite thickness is presented. Using Laplace transform method, the radiation and material energy densities are obtained as a function of space and time. The variation in integrated energy densities and leakage currents are also studied. In order to linearize the radiation transport and material energy equation, the heat capacity is assumed to be proportional to the cube of the material temperature. The steady state energy densities show linear variation along the depth of the planar slab, whereas non-linear dependence is observed for the spherical shell. The analytical energy densities show good agreement with those obtained from finite difference method using small mesh width and time step. The benchmark results obtained in this work can be used to validate and verify non equilibrium radiation diffusion computer codes in both planar and spherical geometry.
\end{abstract}



\section{Introduction}

The time dependent non-equilibrium radiation transport equation is non linearly coupled to the material energy equation\citep{Pomraning-book},\citep{Mihalas-book}. Also the material properties have complex dependence on the independent variables. As a result, the time dependent thermal radiation transport problems are commonly solved numerically. Several numerical methods are in use for this purpose, namely the discrete ordinates \citep{Ghosh}, finite volume \citep{Kim}, Monte Carlo \citep{Fleck}, hybrid stochastic-deterministic \citep{Densmore},\citep{Connolly}, or the approximate methods like the Eddington approximation \citep{Shettle}, heat conduction \citep{Goldstein} or the diffusion approximations \citep{Dai}, \citep{Knoll}, \citep{Ober}. Benchmark results for test problems are necessary to validate and verify the numerical codes \citep{Ensman}. Analytical solutions producing explicit expressions for the radiation and material energy density, integrated densities, leakage currents, etc. are the most desirable.

In the literature, considerable amount of efforts have been applied for solving the Radiation Transport problem analytically. Marshak obtained a semi-analytical solution by considering radiation diffusion in a semi infinite planar slab with radiation incident upon the surface \citep{Marshak-original}. Assuming that the radiation and material fields are in equilibrium, the problem admits a similarity solution to a second order ordinary differential equation which was solved numerically \citep{Kass}. The results were extended for non-equilibrium radiation diffusion by assuming that the specific heat is proportional to the cube of the temperature \citep{Pomraning}, \citep{Su-olson1}. This assumption linearized the problem providing a detailed analytical solution. As the radiative transfer codes are meant to handle an arbitrary temperature dependence of the material properties, the obtained  solutions serve as a useful test problem \citep{Ganapol-pomraning}, \citep{Su-olson2}, \citep{Su-olson3}. Using the same linearization, 3T radiation diffusion equations were solved for spherical and spherical shell sources in an infinite medium \citep{MCClarren}. All available results on the non-equilibrium radiative transfer problems in planar and spherical geometry consider systems having infinite or semi-infinite extension. Benchmarks involving finite size systems have been limited either to the heat conduction or equilibrium diffusion approximation \citep{Williams},  \citep{Olson-H}, \citep{Liemert}. 

In this paper, we solve the time dependent non-equilibrium radiation diffusion problem for finite size systems in both planar and spherical geometry. Non-equilibrium diffusion codes can be more easily validated and verified against these benchmark results because there is no need to consider a slab or spherical medium of  very large size for avoiding boundary effects. Analytical expressions for all the quantities of interest can be obtained for finite slab/shell width and parameter values relevant to practical problems. This work can be extended to multi-dimension using separation of variables and Laplace transform method or the eigenfunction expansion method to obtain analytical series solution in a manner similar to the multilayer heat conduction\citep{Jain}. 

The remainder of the paper is organised as follows. In Section \ref{analytic}, the analytical solution for the finite planar slab and spherical shell is derived followed by Section \ref{numerical} on numerical finite difference method. In Section \ref{results}, the results for the radiation and material energy densities, leakage currents, integrated quantities, etc. are plotted and physically explained. Finally, conclusions are given in Section \ref{conc}. 
\label{intro}
\section{Analytical solution}\label{analytic}
\subsection{Planar slab}\label{planar}
\begin{figure}
\begin{center}
\includegraphics*[width=8cm]{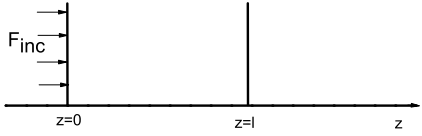}
\end{center}
\caption{\label{slab}Flux incident on the left surface of a slab of thickness $z=l$.}
\end{figure}

We consider a planar slab of finite thickness which is purely absorbing and homogeneous occupying $0\leq z\leq l$. The medium is at zero temperature initially. At time t=0, a constant radiative flux ($F_{inc}$) is incident on the surface at z=0 as shown in Fig. \ref{slab}. Neglecting hydrodynamic motion and heat conduction, the one group radiative transfer equation (RTE) in the diffusion approximation and the material energy balance equation (ME) are \citep{Pomraning-book}
\begin{eqnarray}
\frac{\partial E(z,t)}{\partial t}-\frac{\partial}{\partial z}[\frac{c}{3\kappa (T)}\frac{\partial E(z,t)}{\partial z}]=c \kappa (T) [a T^4 (z,t)-E(z,t)]\label{RTE1}\\ 
C_v(T)\frac{\partial T(z,t)}{\partial t}=c \kappa (T) [E(z,t)-a T^4 (z,t)]\label{ME1}
\end{eqnarray}
where E(z,t) is the radiation energy density, T(z,t) is the material temperature, $\kappa (T)$ is the opacity (absorption cross section), c is the speed of light, $a$ is the radiation constant, and $C_v (T)$ is the specific heat of the material.

The Marshak boundary condition on the surface at $z=0$ is given by
\begin{eqnarray}
E(0,t)-(\frac{2}{3\kappa [T(0,t)]})\frac{\partial E(0,t)}{\partial z}=\frac{4}{c}F_{inc}
\end{eqnarray} 
where $F_{inc}$ is the flux incident upon the surface z=0.

And that at $z=l$ is
\begin{eqnarray}
E(l,t)+(\frac{2}{3\kappa [T(l,t)]})\frac{\partial E(l,t)}{\partial z}=0
\end{eqnarray}
The initial conditions on these two equations are 
\begin{eqnarray}
E(z,0)=T(z,0)=0
\end{eqnarray} 
To remove the nonlinearity in the RTE (Eq.(\ref{RTE1})) and ME (Eq.(\ref{ME1})), opacity $\kappa $ is assumed to be independent of temperature and specific heat $C_v$ is assumed to be proportional to the cube of the temperature. i.e., $C_v=\alpha T^3$.
The RTE and the ME are recast into the dimensionless form by introducing the dimensionless independent variables given by
\begin{eqnarray}
x\equiv \sqrt{3}\kappa z, \tau \equiv (\frac{4ac\kappa }{\alpha })t
\end{eqnarray}
and new dependent variables given by 
\begin{eqnarray}
u(x,\tau)\equiv (\frac{c}{4})[\frac{E(z,t)}{F_{inc}}], v(x,\tau)\equiv (\frac{c}{4})[\frac{a T^4(z,t)}{F_{inc}}]
\end{eqnarray}
With these new variables, the RTE and ME take the dimensionless form 
\begin{eqnarray}\label{RTE2}
\epsilon \frac{\partial u(x,\tau)}{\partial \tau }=\frac{\partial ^2 u(x,\tau)}{\partial x^2}+v(x,\tau)-u(x,\tau) \\ 
\frac{\partial v(x,\tau)}{\partial \tau}=u(x,\tau)-v(x,\tau)\label{ME2}
\end{eqnarray}
with the initial conditions 
\begin{eqnarray}
u(x,0)=0\\\label{u-initial}
v(x,0)=0\label{v-initial}
\end{eqnarray}
And the boundary conditions on the surfaces are
\begin{eqnarray}\label{BC-0}
u(0,\tau)-\frac{2}{\sqrt{3}}\frac{\partial u(0,\tau)}{\partial x}=1\\
u(b,\tau)+\frac{2}{\sqrt{3}}\frac{\partial u(b,\tau)}{\partial x}=0\label{BC-b}
\end{eqnarray}
where $b=\sqrt{3}\kappa l$
and the parameter $\varepsilon $ is defined as 
\begin{eqnarray}
\varepsilon =\frac{16 \sigma }{c \alpha }=\frac{4a}{\alpha }
\end{eqnarray}
To solve Eqs. (\ref{RTE2}) - (\ref{BC-b}), we introduce the Laplace transform according to 
\begin{eqnarray}
\bar f (s)=\int_0^\infty d\tau e^{-s\tau}f(\tau)
\end{eqnarray}
to obtain
\begin{eqnarray}
\varepsilon s \bar u (x,s)-\frac{\partial^2 \bar u(x,s)}{\partial x^2}=\bar v(x,s)-\bar u(x,s)\label{RTE-s}\\
s\bar v(x,s)=\bar u(x,s)-\bar v(x,s)\\\label{ME-s}
\bar u(0,s)-\frac{2}{\sqrt{3}}\frac{\partial \bar u(0,s)}{\partial x}=\frac{1}{s}\\\label{BC-0-s}
\bar u(b,s)+\frac{2}{\sqrt{3}}\frac{\partial \bar u(b,s)}{\partial x}=0\label{BC-a-s}
\end{eqnarray}
The solutions of Eqs. (\ref{RTE-s})-(\ref{BC-a-s}) in s space are obtained as 
\begin{eqnarray}
\bar u(x,s)=\frac{3sin[\beta(s)(b-x)]+2\sqrt{3}\beta(s) cos[\beta(s)(b-x)]}{s[3sin(\beta(s) b)+4\sqrt{3}\beta(s) cos(\beta(s) b)-4 \beta^2(s)  sin(\beta(s) b)]}\label{u-bar}\\
\bar v(x,s)=\frac{3sin[\beta(s)(b-x)]+2\sqrt{3}\beta(s) cos[\beta(s)(b-x)]}{s(s+1)[3sin(\beta(s) b)+4\sqrt{3}\beta(s) cos(\beta(s) b)-4 \beta^2(s) sin(\beta(s) b)]}\label{v-bar}
\end{eqnarray}
where $\beta (s)$ is given by 
\begin{eqnarray}
\beta^2(s)=-\frac{s}{s+1}[1+\varepsilon (s+1)]\label{beta-square}
\end{eqnarray}

Before solving for the radiation and material energy densities by inverting $\bar u(x,s)$ and $\bar v(x,s)$, we first obtain the small and large $\tau$ limits of $u(x,\tau)$ and $v(x,\tau)$ from the large and small s limits of Eqs. (\ref{u-bar}) and (\ref{v-bar}) respectively. Using the theorems 
\begin{eqnarray}
lim_{s \rightarrow \infty}[s\bar f(s)]=lim_{\tau \rightarrow 0}[f(\tau)]\\
lim_{s \rightarrow 0}[s\bar f(s)]=lim_{\tau \rightarrow \infty}[f(\tau)]
\end{eqnarray} 
we have
\begin{eqnarray}
u(x,0)=v(x,0)=0\label{0-time}\\
u(x,\tau \rightarrow \infty)\rightarrow v(x,\tau \rightarrow \infty)\rightarrow \frac{3b+2\sqrt{3}-3x}{3b+4\sqrt{3}}\label{infinite-time}
\end{eqnarray}
Thus according to Eq. (\ref{0-time}), at the initial instant, both the material and radiation energy densities are zero inside the slab. Eq. (\ref{infinite-time}) asserts that at infinite time the radiation and material energy density equilibrate among themselves. However, because of the finite thickness of the slab, flux leaks out of the right edge so that the energy densities vary linearly along the length of the slab.

\begin{figure}
\begin{center}
\includegraphics*[width=8cm]{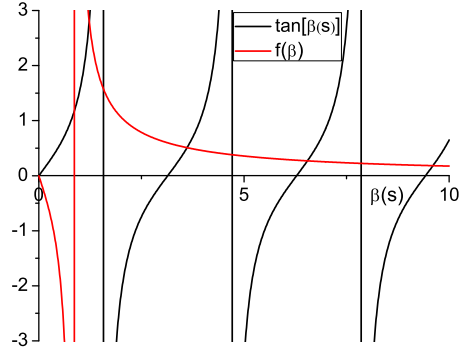}
\end{center}
\caption{\label{root}Finding the roots of the transcendental equation$\ tan(\beta(s) )=f(\beta)=\frac{4\sqrt{3}\beta(s)}{4\beta^2(s)-3}$ .}
\end{figure}

The solutions for $ u(x,\tau)$ and  $v(x,\tau)$ follow from $\bar u (x,s)$ and $\bar v (x,s)$ by inverting them using the Laplace inversion theorem
\begin{eqnarray}
f(\tau)=\frac{1}{2\pi i}\int_C ds e^{s\tau} \bar f(s)
\end{eqnarray}
where the integration contour is a line parallel to the imaginary s axis to the right of all the singularities of $\bar f(s)$.The contour is closed in the left half plane so that the large semi circle gives a zero contribution. Both $\bar u(x,s)$ and $\bar v(x,s)$ are single valued functions and hence there are no branch points. However, there are an infinite number of poles obtained from the roots of the transcendental equation 
\begin{eqnarray}
3sin(\beta(s) b)+4\sqrt{3}\beta(s) cos(\beta(s) b)-4 \beta ^2(s) sin(\beta(s) b)=0\\
or, tan(\beta(s) b)=\frac{4\sqrt{3}\beta(s)}{4\beta^2(s)-3}
\end{eqnarray}
For the semi infinite slab, because of the multiple valuedness of the functions obtained by Laplace transform, inverting them using the inverse Laplace transform required evaluation of contributions from all the branch cuts. This resulted in integrals which had to be computed numerically \citep{Su-olson1}. The oscillations in the integrand resulted in difficulty in their convergence. The advantage of solving the finite problem is that because of the single valuedness of the Laplace transformed functions, the inversion is very simple. The sum of the residues at the singularities (poles) give the required solution.
The roots of the transcendental equation has been obtained using MATHEMATICA \citep{mathematica} as shown in the graph of Fig. \ref{root}.

Corresponding to each root of $\beta(s)$, there exists two values of s, i.e., two simple poles. The poles are obtained from solution of Eq. (\ref{beta-square})as $\frac{-(\epsilon+\beta^2(s)+1)\pm \sqrt{(\epsilon+\beta^2(s)+1)^2-4\epsilon \beta^2(s)})}{2 \epsilon}$. According to the residue theorem, $\int_C ds e^{s\tau} \bar f(s)=2 \pi i \times  $(sum of the residues at the singularities).
The residue at s=0 gives the asymptotic (steady state) solution for the radiation and material energy densities as  $u(x,\infty)=v(x,\infty)=\frac{3b+2\sqrt{3}-3x}{3b+4\sqrt{3}}$ which is also obtained by equating $\frac{\partial u(x,\tau)}{\partial \tau}$ and $\frac{\partial v(x,\tau)}{\partial \tau}$ in Eqs. (\ref{RTE2}) and (\ref{ME2}) to zero, solving $\frac{\partial^2 u(x,\tau)}{\partial x^2}=0$ and obtaining the values of the constants from the BC given by Eqs. (\ref{BC-0}) and (\ref{BC-b}).

The contribution to the time dependent part comes from the higher order poles. Adding residues from all the poles give us the complete space and time dependence of the radiation energy density as 
\begin{eqnarray}
u(x,\tau)=\frac{3b+2\sqrt{3}-3x}{3b+4\sqrt{3}}\nonumber
\end{eqnarray}
\begin{eqnarray}
+\sum_{n} \frac{e^{s_n \tau}[3sin(\beta(s_n)(b-x))+2\sqrt{3}\beta(s_n) cos(\beta(s_n) (b-x))]}{s_n[(3b+4\sqrt{3}-4\beta^2(s_n) b)cos(\beta(s_n) b)-(4\sqrt{3}\beta(s_n) b+8\beta(s_n)) sin(\beta(s_n) b)]\frac{d\beta(s_n)}{ds}}\nonumber\\ 
\end{eqnarray}

Similarly, the solution for the material energy density is 
\begin{eqnarray}
v(x,\tau)=\frac{3b+2\sqrt{3}-3x}{3b+4\sqrt{3}}\nonumber
\end{eqnarray}
\begin{eqnarray}
+\sum_{n} \frac{e^{s_n \tau}[3sin(\beta(s_n)(b-x))+2\sqrt{3}\beta(s_n) cos{\beta(s_n) (b-x)}]}{s_n(s_n +1)[(3b+4\sqrt{3}-4\beta^2(s_n) b)cos(\beta(s_n) b)-(4\sqrt{3}\beta(s_n) b+8\beta(s_n)) sin(\beta(s_n) b)]\frac{d\beta(s_n)}{ds}}\nonumber\\
\end{eqnarray}

We also consider the $\epsilon $=0 case which arises when the speed of light is taken to be infinite so that radiation is not retarded initially. At infinite time, the radiation and material energy densities assume the same spatial dependence as for $\epsilon \neq 0$ case.
\begin{eqnarray} 
u(x,\tau \rightarrow \infty) \rightarrow v(x,\tau \rightarrow \infty)\rightarrow \frac{3b+2\sqrt{3}-3x}{3b+4\sqrt{3}}\label{infinite-time-eps0}
\end{eqnarray}
However, for $\tau=0$, as $s \rightarrow \infty $ for $\epsilon=0$, we obtain $\beta=i$ where $i=\sqrt{-1}$. Thus, 
\begin{eqnarray}
u(x,0)=\frac{3sinh(b-x)+2\sqrt{3}cosh(b-x)}{7 sinh(b)+4\sqrt{3}cosh(b)}\\
v(x,0)=0
\end{eqnarray}
Thus the material energy density is zero at $\tau=0$ as predicted by the initial condition. However, because of the absence of retardation effects, the radiation energy density attains a finite value consistent with the incoming flux of radiation. This behaviour is in agreement with that obtained in the case of a semi infinite planar slab for the no retardation case. 

The solution $u(x,\tau)$ and $v(x,\tau)$ for $\epsilon=0 $ is obtained by inverting Eqs. ($\ref{u-bar}$) and ($\ref{v-bar}$) using inverse Laplace transform as in the general case with $\epsilon = 0$. The difference from the $\epsilon \ne 0$ case is that only one pole is obtained corresponding to a value of beta i.e., $s=-\frac{\beta^2(s)}{\beta^2(s)+1}$.

\subsection{Spherical shell}\label{spherical}
\begin{figure}
\begin{center}
\includegraphics*[width=8cm]{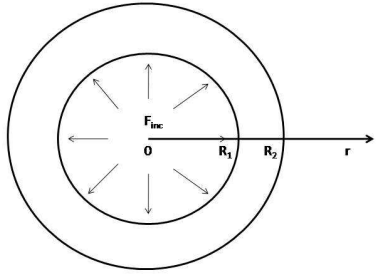}
\end{center}
\caption{\label{shell}Flux incident on the inner surface of a spherical shell of inner radius $R_1$ and outer radius $R_2$.}
\end{figure}

Analogous to the planar slab problem, in spherical geometry we consider a spherical shell of inner and outer radii $R_1$ and $R_2$ respectively (Fig. \ref{shell}). Under the same assumptions, with a time independent radiative flux ($F_{inc}$) incident on the inner surface of the shell, the one group radiative transfer equation (RTE) in the diffusion approximation and the material energy balance equation (ME) in spherical geometry are 
\begin{eqnarray}
\frac{\partial E(r,t)}{\partial t}-\frac{1}{r^2}\frac{\partial}{\partial r}[\frac{r^2 c}{3\kappa (T)}\frac{\partial E(r,t)}{\partial r}]=c \kappa (T) [a T^4 (r,t)-E(r,t)]\label{RTE1s}\\ 
C_v(T)\frac{\partial T(r,t)}{\partial t}=c \kappa (T) [E(r,t)-a T^4 (r,t)]\label{ME1s}
\end{eqnarray}
with the same notations as used in Subsec. \ref{planar}

The Marshak boundary condition on the inner surface at $r=R_1$ is given by
\begin{eqnarray}
E(R_1,t)-(\frac{2}{3\kappa [T(R_1,t)]})\frac{\partial E(R_1,t)}{\partial r}=\frac{4}{c}F_{inc}
\end{eqnarray} 

And that at $r=R_2$ is
\begin{eqnarray}
E(R_2,t)+(\frac{2}{3\kappa [T(R_2,t)]})\frac{\partial E(R_2,t)}{\partial r}=0
\end{eqnarray}

With new dimensionless variables introduced in Subsec. \ref{planar}, the RTE and ME take the dimensionless form 
\begin{eqnarray}\label{RTE2s}
\epsilon \frac{\partial u(x,\tau)}{\partial \tau }=\frac{1}{x^2}\frac{\partial}{\partial x}(x^2 \frac{\partial u(x,\tau)}{\partial x})+v(x,\tau)-u(x,\tau) \\ 
\frac{\partial v(x,\tau)}{\partial \tau}=u(x,\tau)-v(x,\tau)\label{ME2s}
\end{eqnarray}
with the initial conditions 
\begin{eqnarray}
u(x,0)=0\\\label{u-initial}
v(x,0)=0\label{v-initial}
\end{eqnarray}
And the boundary conditions on the surfaces are
\begin{eqnarray}
u(X_1,\tau)-\frac{2}{\sqrt{3}}\frac{\partial u(X_1,\tau)}{\partial x}=1\label{BC-x1}\\
u(X_2,\tau)+\frac{2}{\sqrt{3}}\frac{\partial u(X_2,\tau)}{\partial x}=0\label{BC-x2}
\end{eqnarray}
where $x=\sqrt{3}\kappa r$

Changing variable $u(x,\tau)$ to $w(x,\tau)=u(x,\tau)x$ and $v(x,\tau)$ to $g(x,\tau)=v(x,\tau)x$, 
the equations simplify to 
\begin{eqnarray}
\epsilon \frac{\partial w(x,\tau)}{\partial \tau }=\frac{\partial ^2 w(x,\tau)}{\partial x^2}+g(x,\tau)-w(x,\tau) \\ 
\frac{\partial g(x,\tau)}{\partial \tau}=w(x,\tau)-g(x,\tau)
\end{eqnarray}

Applying Laplace transform, the solution in s space are obtained as 
\begin{eqnarray}
\bar u(x,s)=\frac{A}{\beta(s) x}sin(\beta(s) x+B)\\
\bar v(x,s)=\frac{\bar u(x,s)}{s+1}
\end{eqnarray}
with the constants A and B obtained from the BCs
\begin{eqnarray}
\bar u(X_1,s)-\frac{2}{\sqrt 3}\frac{\partial \bar u(X_1,s)}{\partial x}=\frac{1}{s}\\
\bar u(X_2,s)+\frac{2}{\sqrt 3}\frac{\partial \bar u(X_2,s)}{\partial x}=0
\end{eqnarray} 
Then the Laplace transformed radiation energy density is given by

 $\bar u = \frac{\sqrt{3} X_1^2[(2-\sqrt{3}X_2)sin(\beta(X_2-x))-2\beta X_2 cos(\beta(X_2-x))]}{s x [((4\beta^2-3)X_1X_2-2\sqrt{3}(X_2-X_1)+4)sin{\beta(X_2-X_1)}-(4\beta (X_2-X_1)+4\sqrt{3}\beta X_1 X_2)cos(\beta(X_2-X_1))]}$

As in the case of the finite planar slab, the solutions for $ u(x,\tau)$ and  $v(x,\tau)$ follow from $\bar u (x,s)$ and $\bar v (x,s)$ by inverting them using the Laplace inversion theorem. An infinite number of poles are obtained from the roots of the transcendental equation 
\begin{eqnarray}
tan(\beta(s)(X_2-X_1))=\frac{4\sqrt{3}\beta(s) X_2 X_1+4 \beta(s) (X_2-X_1)}{(4 \beta(s)^2-3) X_1 X_2-2 \sqrt{3} (X_2-X_1)+4 }
\end{eqnarray}
Summing over the residues at all the poles, the radiation energy density is obtained as 
$ u(x,\tau)=\frac{\sqrt{3} X_1^2 X_2^2 + X_1^2 x (2-\sqrt{3}X_2)}{x[2 X_1^2-\sqrt{3}X_1^2 X_2+\sqrt{3} X_1 X_2^2+2 X_2^2]}\\ 
+\sum_{n}\frac{[(2-\sqrt{3}X_2)sin(\beta(X_2-x))-2\beta X_2 cos(\beta(X_2-x))]}{[((4\beta^2(X_2^2+X_1^2)+4\sqrt{3}\beta X_1 X_2 (X_2-X_1))sin{\beta(X_2-X_1)}+(4\beta^2 X_1 X_2 (X_2-X_1)-3 X_1 X_2 (X_2-X_1)-2\sqrt{3}(X_1^2+X_2^2))cos(\beta(X_2-X_1))]}\\ \times \frac{e^{s_n \tau}\sqrt{3} X_1^2}{s_n x\frac{d\beta(s_n)}{ds}}$.

For convenience in writing the expressions, $\beta(s_n)$ has sometimes been written as $\beta$.
Similarly, the solution for the material energy density follows the same form as that for the radiation energy density with an extra $(s_n+1)$ in the denominator of the second term.

\section{Numerical Finite difference solution}\label{numerical}
In this section, we present the finite difference method for obtaining the energy densities for finite slab and spherical shell numerically. 
\subsection{Planar slab}
We assume that the opacity is temperature independent and the heat capacity is proportional to the cube of the temperature, $C_V=\alpha T^3(z,t)$. Then, for a material energy density  $\theta= a T^4(z,t)$ and radiation flux $F(z,t)=-\frac{c}{3\kappa}\frac{\partial E(z,t)}{\partial z}$, the radiation and material energy density equations along with the boundary conditions for a finite slab of thickness $l$ are
\begin{eqnarray}
\frac{\partial E(z,t)}{\partial t}+\frac{\partial F(z,t)}{\partial z}=c\kappa(\theta (z,t)-E(z,t))\\
\frac{1}{c}\frac{\partial \theta (z,t)}{\partial t}=\epsilon\kappa(E(z,t)-\theta (z,t))\\
cE(0,t)+2F(0,t)=4F_{inc}\\
cE(l,t)-2F(l,t)=0
\end{eqnarray}
Time differencing is performed using a fully implicit backward Euler scheme. Spatial discretization is performed on a staggered mesh where the independent spatial variable z and the flux F are evaluated at cell edges and the energy densities represent cell averages at the cell centers. The finite difference equations for the radiation and material energy densities are obtained as \citep{Su-olson1}
\begin{eqnarray}
(1+\frac{\kappa}{\gamma+\epsilon \kappa})\gamma E_i^{n+1}+\frac{1}{c\Delta z_i}(F_{i+1/2}^{n+1}-F_{i-1/2}^{n+1})=\gamma E_i^n+\frac{\kappa \gamma}{\gamma+\epsilon \kappa}\theta_i^n\\
\theta_i^{n+1}=\frac{\gamma}{\gamma+\epsilon \kappa}\theta_i^n+\frac{\epsilon \kappa}{\gamma+\epsilon \kappa}E_i^{n+1}
\end{eqnarray}
where $\gamma=1/c\Delta t$ and $\Delta z_i = z_{i+1/2}-z_{i-1/2}$.
  The energy density is assumed to be a piecewise linear function in space and we define two fluxes at the cell edge, one from the left and one from the right.
\begin{eqnarray}
F_{l,i+1/2}^{n+1}=-\frac{2c}{3\kappa}\frac{E_{i+1/2}^{n+1}-E_i^{n+1}}{\Delta z_i}\\
F_{r,i+1/2}^{n+1}=-\frac{2c}{3\kappa}\frac{E_{i+1}^{n+1}-E_{i+1/2}^{n+1}}{\Delta z_{i+1}}
\end{eqnarray}
The edge value of the radiation energy density is a weighted average of the cell center quantities. 
Finally a tridiagonal system of equations is obtained for the radiation energy density at time n+1 as 
\begin{eqnarray}
-E_{i-1}^{n+1}+[1+\frac{\Delta z_{i-1/2}}{\Delta z_{i+1/2}}+3\kappa \Delta z_i \Delta z_{i-1/2}\gamma (1+\kappa/(\gamma+\epsilon \kappa))]E_i^{n+1}\nonumber\\-\frac{\Delta z_{i-1/2}}{\Delta z_{i+1/2}}E_{i+1}^{n+1}=3\kappa\Delta z_i\Delta z_{i-1/2}\gamma E_i^n+\frac{3\kappa^2\Delta z_i \Delta z_{i-1/2}\gamma}{\gamma+\epsilon \kappa}\theta_i^n
\end{eqnarray}
where $\Delta z_{i+1/2}=\frac{1}{2}(\Delta z_i+\Delta z_{i+1})$.
Applying the BCs for the first and last cell, the radiation energy density equations for the first cell is 
\begin{eqnarray}
[1+2(\frac{\Delta z_1}{\Delta z_{3/2}}+\frac{4}{3\kappa \Delta z_{3/2}})^{-1}+3\kappa \Delta z_1 \Delta z_{3/2} \gamma (1+\frac{\kappa}{\gamma+\epsilon \kappa})]E_1^{n+1}-E_2^{n+1}\nonumber\\=3\kappa \Delta z_1 \Delta z_{3/2}\gamma (E_1^n+\frac{\kappa}{\gamma+\epsilon \kappa}\theta_1^n)+\frac{8}{c}F_{inc}(\frac{\Delta z_1}{\Delta z_{3/2}}+\frac{4}{3\kappa \Delta z_{3/2}})^{-1} 
\end{eqnarray}
And that for the last cell is 
\begin{eqnarray}
[1+(\frac{\Delta z_N}{\Delta z_{N-1/2}}+\frac{4}{3\kappa \Delta z_{N-1/2}})^{-1}+3\kappa \Delta z_N \Delta z_{N-1/2} \gamma (1+\frac{\kappa}{\gamma+\epsilon \kappa})]E_N^{n+1}\nonumber\\-E_{N-1}^{n+1}=3\kappa \Delta z_N \Delta z_{N-1/2}\gamma (E_N^n+\frac{\kappa}{\gamma+\epsilon \kappa}\theta_N^n)
\end{eqnarray}

\subsection{Spherical shell}
In this section, we derive the finite difference equations for obtaining the radiation and material energy densities for a spherical shell of inner radius $R_1$ and outer radius $R_2$. Using the transformation $E^\prime(r,t)=E(r,t)r$ and $\theta^\prime=aT^4(r,t)r$, the flux is defined as $F(r,t)=-\frac{c}{3\kappa}\frac{\partial E^\prime(r,t)}{\partial r}$.
Then the equations for transformed radiation and material energy densities and the boundary conditions are
\begin{eqnarray}
\frac{\partial E^\prime (r,t)}{\partial t}+\frac{\partial F(r,t)}{\partial r}=c\kappa(\theta^\prime (r,t)- E^\prime (r,t))\\
\frac{1}{c}\frac{\partial \theta^\prime (r,t)}{\partial t}=\epsilon\kappa( E^\prime (r,t)-\theta^\prime (r,t))\\
(\frac{1}{R_1}+\frac{2}{3\kappa R_1^2}) E^\prime (R_1,t)+\frac{2}{c}\frac{F(R_1,t)}{R_1}=\frac{4F_{inc}}{c}\\
(\frac{1}{R_2}-\frac{2}{3\kappa R_2^2}) E^\prime (R_1,t)-\frac{2}{c}\frac{F(R_2,t)}{R_2}=0
\end{eqnarray}
Using finite differencing in space and time as done for the plane slab, the tridiagonal equation for energy density (in terms of the transformed variables $ E^\prime$ and $\theta^\prime$) of the inner cells is 
\begin{eqnarray}
-{E^\prime}_{i-1}^{n+1}+[1+\frac{\Delta r_{i-1/2}}{\Delta r_{i+1/2}}+3\kappa \Delta r_i \Delta r_{i-1/2}\gamma (1+\frac{\kappa}{\gamma+\epsilon \kappa})]{E^\prime}_i^{n+1}\nonumber\\-\frac{\Delta r_{i-1/2}}{\Delta r_{i+1/2}}{E^\prime}_{i+1}^{n+1}=3 \kappa \Delta r_i \Delta r_{i-1/2} \gamma {E^\prime}_i^n+\frac{3 \kappa^2 \Delta r_i \Delta r_{i-1/2} \gamma}{\gamma+\epsilon \kappa}\theta_i^{\prime n}
\end{eqnarray}
From the BC on the surface of the first and last cell, we obtain the equation relating the energy densities for the first cell as
\begin{eqnarray}
[1+2 \frac{(2+3 \kappa R_1)\Delta r_{3/2}}{4R_1+3 \kappa R_1 \Delta r_1+2\Delta r_1}+3 \kappa \Delta r_1 \Delta r_{3/2} \gamma (1+\frac{\kappa}{\gamma+\epsilon \kappa})]{E^\prime}_1^{n+1}-{E^\prime}_2^{n+1}\nonumber\\=3\kappa\Delta r_1 \Delta r_{3/2}\gamma({E^\prime}_1^n+\frac{\kappa \theta_1^{\prime n}}{\gamma+\epsilon \kappa})+\frac{24F_{inc}}{c}\frac{1}{\frac{4}{R_1\kappa\Delta r_{3/2}}+\frac{3\Delta r_1}{\Delta r_{3/2}R_1}+\frac{2\Delta r_1}{R_1^2 \kappa \Delta r_{3/2}}}]
\end{eqnarray}
Similarly, the equation for the last cell is
\begin{eqnarray}
[1+\frac{2\Delta r_{N-1/2}(3\kappa R_2-2)}{(3\kappa R_2-2)\Delta r_N+4R_2}+3\kappa \Delta r_N \Delta r_{N-1/2}\gamma (1+\frac{\kappa}{\gamma+\epsilon \kappa})]{E^\prime}_N^{n+1}\nonumber\\-{E^\prime}_{N-1}^{n+1}=3\kappa\Delta r_ \Delta r_{N-1/2}\gamma({E^\prime}_N^n+\frac{\kappa \theta_N^{\prime n}}{\gamma+\epsilon \kappa})
\end{eqnarray}

\section{Results and discussions}\label{results}

\subsection{Planar slab}
\begin{figure}
\begin{center}
\includegraphics*[width=8cm]{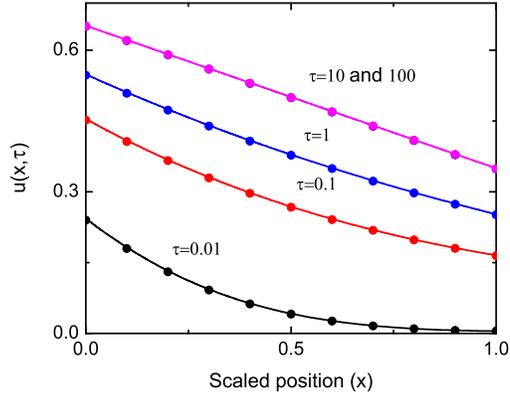}
\end{center}
\caption{\label{uxt-planar}Scaled radiation energy density $u(x,\tau)$ vs position (x) in the slab of scaled thickness $b=1$ at different times for $ \epsilon = 0.1 $. The symbols stand for analytical values whereas lines represent the results obtained from finite difference method.}
\end{figure}

\begin{figure}
\begin{center}
\includegraphics*[width=8cm]{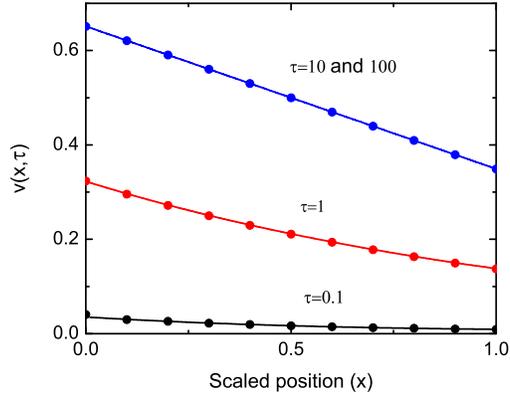}
\end{center}
\caption{\label{vxt-planar}Scaled material energy density $v(x,\tau)$ vs position (x) in the slab at different times for $\epsilon = 0.1$. The symbols stand for analytical values whereas lines represent the results obtained from finite difference method.}
\end{figure}

The radiation and material energy densities  obtained from finite difference analysis are plotted in Figs. \ref{uxt-planar} and \ref{vxt-planar} along with the analytical results for a slab of width $b=1$. To obtain a normalized solution that is comparable to analytical solution, we choose $F_{inc}=c/4$ , so that E and $\theta $ directly correspond to u and v. The numerical results obtained from finite difference analysis are found to converge for a mesh width $\Delta z=5.7733 \times 10^{-5}$ cm. A time step of $\Delta t = 3.33 \times 10^{-15}$ s is chosen upto a scaled time $\tau = 0.1$. Beyond this time, a coarser time step of $\Delta t = 3.33 \times 10^{-12}$ s is found to be sufficient for obtaining the converged values. The numerical results are found to agree with the analytical ones with an error $< 1 \%$ at early stages ($\tau = 0.01$). The error reduces further as time progresses. The close agreement between analytical and numerical results prove the correctness of both the methods. For the finite planar slab, at early stages ($\tau$=0.01) the radiation energy density falls rapidly from the left surface where radiation is incident as shown in Fig. \ref{uxt-planar}. As time proceeds, the values of energy densities increase and the variation with distance keeps on attaining linearity. At infinite time, the steady state values are linear with position as given by Eq. \ref{infinite-time}. Similarly, the material energy density initially exhibits slight non-linear variation and finally attains the linearity (Fig. \ref{vxt-planar}). The non-linear variation at early stages occurs due to net absorption of energy by the initially cold material (as u(x,0)=v(x,0)=0). Initially, the material energy density is found to lag behind the radiation energy densities and finally equilibrate as time proceeds (beyond $\tau$=10). In this work, all the results have been obtained by considering contribution from the first 30 roots of the transcendental equation. The value of opacity $\kappa$ is chosen to be 100 and $\epsilon $ equals 0.1.  For a heat wave traveling into a thin plate and composite planar slab, a similar linear variation in temperature with distance was observed though difference existed in the space and time dependent behaviour due to heat conduction approximation \citep{Sarkar},\citep{Sun}. 

\begin{figure}
\begin{center}
\includegraphics*[width=8cm]{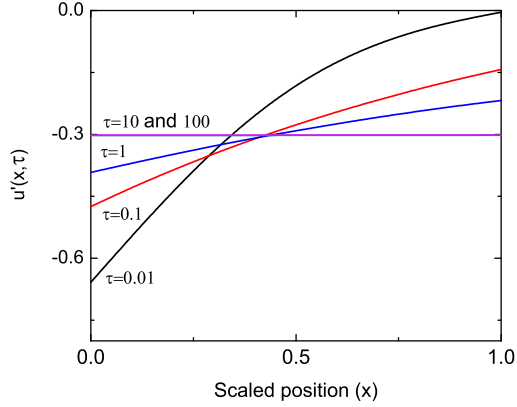}
\end{center}
\caption{\label{uprimext-planar}Space derivative of scaled radiation energy density $ u^\prime(x,\tau)$ vs position (x) in the slab at different times.}
\end{figure}

\begin{figure}
\begin{center}
\includegraphics*[width=8cm]{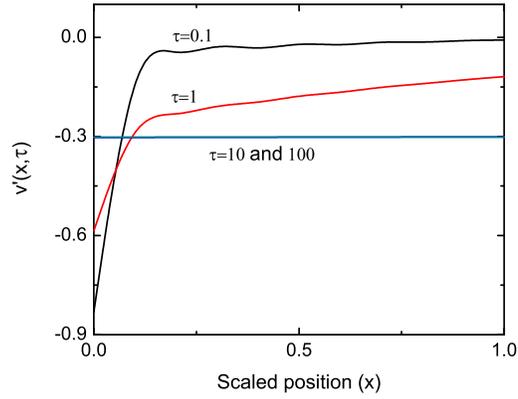}
\end{center}
\caption{\label{vprimext-planar}Space derivative of scaled material energy density $ v^\prime(x,\tau)$ vs position (x) in the slab at different times.}
\end{figure}

The first derivatives w.r.t. position of the analytical radiation and material energy density are plotted in Figs. \ref{uprimext-planar} and \ref{vprimext-planar}. As the radiation and material energy densities decrease with x, the derivative has negative values. The derivative has a greater negative value at the left compared to the right zone. As both radiation and material energy densities keep on increasing with time due to radiation diffusion, magnitude of the gradient decreases for the left and increases for the right sides. The gradient of both radiation and material energy densities obtain a constant value of $\frac{-3}{3+4\sqrt{3}}=-0.30217$ after infinite time showing that there is a constant leakage of flux from the right surface due to the finite thickness. This result is different from the semi-infinite slab result where at infinite time, the entire halfspace is at a constant temperature with a uniform radiation field and hence there is no gradient and no flux \citep{Pomraning}.  

\begin{figure}
\begin{center}
\includegraphics*[width=8cm]{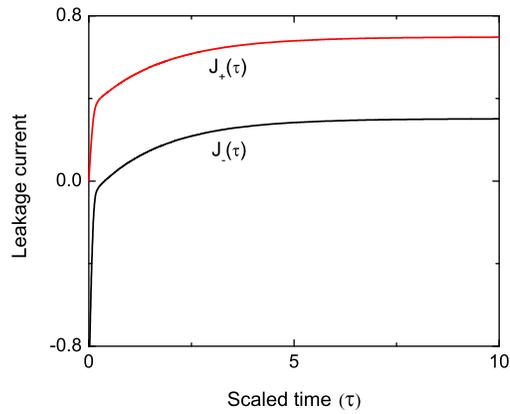}
\end{center}
\caption{\label{jplusminus-planar}Leakage currents $J_-(\tau)$ and $J_+(\tau)$ from the left and right surfaces of the slab respectively.}
\end{figure}

The current of radiation leaking out from the left and right surfaces of the slab are $ J_-(\tau)= u(0,\tau)+\frac{2}{\sqrt{3}}\frac{\partial u(0,\tau)}{\partial x}$ and $J_+(\tau) = u(b,\tau)-\frac{2}{\sqrt{3}}\frac{\partial u(b,\tau)}{\partial x}$.
The leakage currents are plotted as a function of time in Fig. \ref{jplusminus-planar}. It is found that though $J_-(\tau)$ is negative initially, it attains a constant positive value of 0.30217 after saturation. $J_+(\tau)$ is zero initially as the incident flux has not reached the right face. However it rises rapidly and reaches a saturation value of 0.6978. The energy densities and leakage currents at the left and right surfaces are also related as $u(0,\tau)+u(b,\tau)=1$ and $J_-(\tau)+J_+(\tau)=1$.  

\begin{figure}
\begin{center}
\includegraphics*[width=8cm]{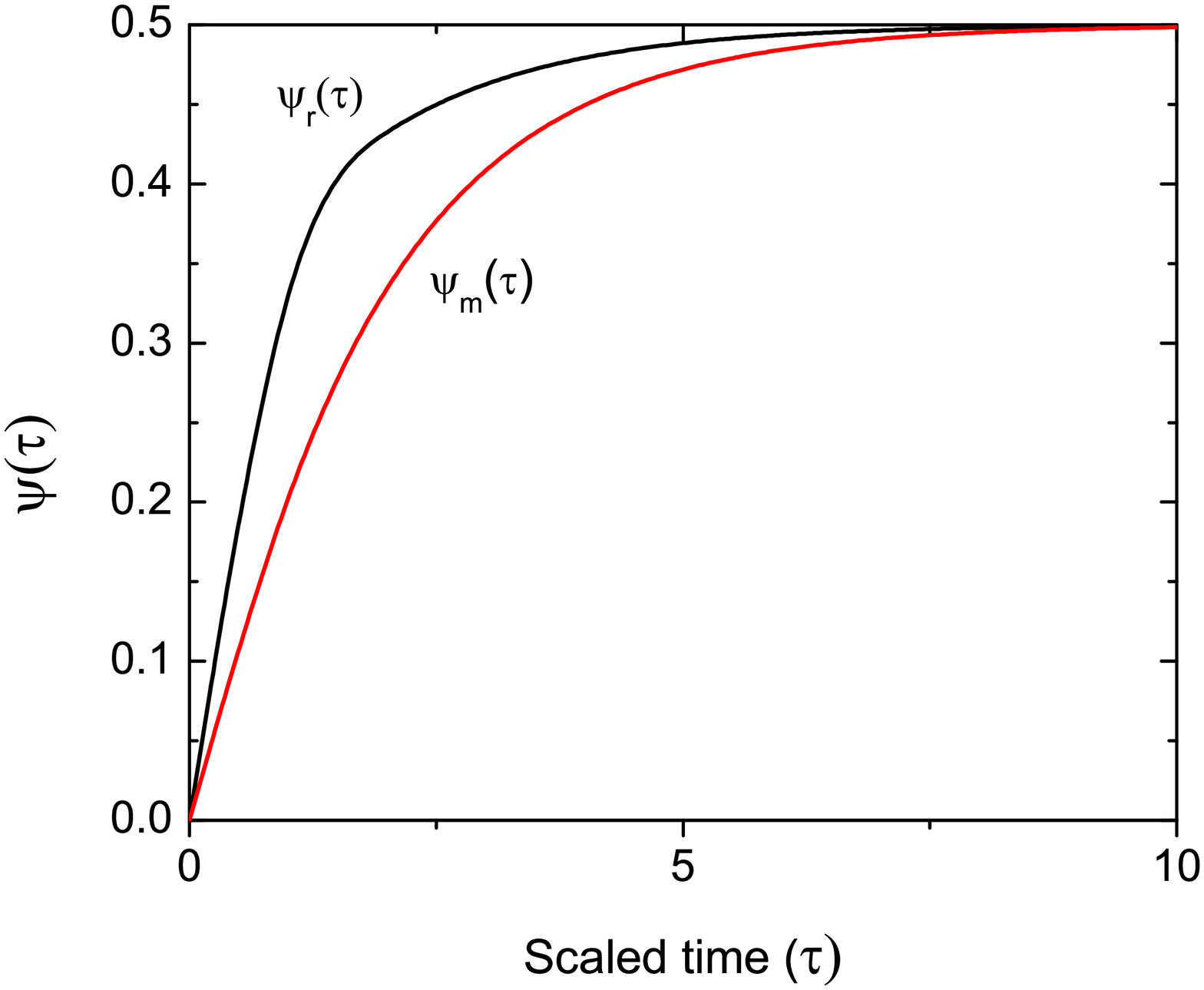}
\end{center}
\caption{\label{average-planar}Integrated radiation ($\psi_r(\tau)$) and material energy densities ($\psi_m(\tau)$) in the slab as a function of  scaled time $\tau$.}
\end{figure}

The averaged or integrated radiation and material energy densities are given by 
$\psi_r(\tau)=\int_0^b u(x,\tau)dx$ and $\psi_m(\tau)=\int_0^b v(x,\tau)dx$ respectively. The steady state integrated value is 0.5 as seen from Fig.\ref{average-planar}. The integrated material energy density is also found to lag the radiation energy density at early times but finally the two equilibrate.

To check the consistency of the final results, we add Eqs. (\ref{RTE2}) and (\ref{ME2}) and integrate over x from 0 to b, yielding 
\begin{eqnarray}
\int_0^b(\epsilon \frac{\partial u(x,\tau)}{\partial \tau}+\frac{\partial v(x,\tau)}{\partial \tau})dx=\int_0^b \frac{\partial ^2 u(x,\tau)}{\partial x^2}dx= \frac{\partial u(b,\tau)}{\partial x}-\frac{\partial u(0,\tau)}{\partial x}\nonumber
\end{eqnarray}
i.e.,
\begin{eqnarray}
\epsilon \frac{\partial \psi_r(\tau)}{\partial \tau}+\frac{\partial \psi_m(\tau)}{\partial \tau}=\frac{\partial u(b,\tau)}{\partial x}-\frac{\partial u(0,\tau)}{\partial x}
\end{eqnarray}
Using the expressions for the energy densities, their first derivatives in space and the integrated quantities, we find that both the left and right hand sides reduce to $\sum_{n} \frac{e^{s_n \tau}}{[(3b+4\sqrt{3}-4\beta^2(s_n) b)cos(\beta(s_n) b)-(4\sqrt{3}\beta(s_n) b+8\beta(s_n)) sin(\beta(s_n) b)]\frac{d\beta(s_n)}{ds}}
\times [\frac{3}{\beta(s_n)}(1-cos(\beta(s_n) b))+2\sqrt{3}sin(\beta(s_n) b)](\epsilon +\frac{1}{s_n +1})$ proving the consistency of the obtained solutions.

\begin{figure}
\begin{center}
\includegraphics*[width=8cm]{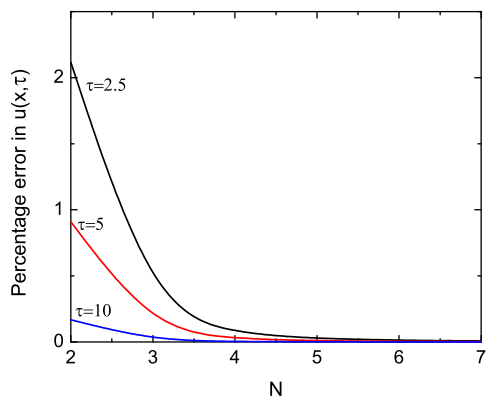}
\end{center}
\caption{\label{convergence-planar}Percentage error in the radiation energy density $u(x,\tau)$ in the slab as a function of number of roots considered (N).}
\end{figure}

As there are infinite number of residues, the exact solution is obtained only on adding all of them. However, the contribution from the poles decrease very sharply. To study convergence, we plot percentage error as a function of number of roots of the transcendental equation considered. As seen from Fig. \ref{convergence-planar}, 2.1 $\%$ error in the value of $u(0, 2.5)$ is observed on considering only the first two roots i.e., the steady state result and residue for the two non zero  poles. The errors arising due to non inclusion of higher order terms is more initially as the higher order poles contribute only at very small times because of the exponential term. The error falls sharply to a negligible value (0.005$\%$) on considering the contribution from the first 6 roots i.e., first 11 poles. More accurate results can be obtained by adding residues from higher order poles.

\begin{figure}
\begin{center}
\includegraphics*[width=8cm]{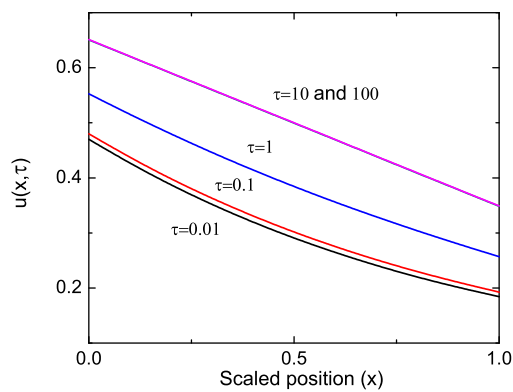}
\end{center}
\caption{\label{uxt-planar-eps0}Scaled radiation energy density $u(x,\tau)$ vs position (x) in the slab of scaled thickness $b=1$ at different times for $ \epsilon = 0$.}
\end{figure}

Fig. \ref{uxt-planar-eps0} shows the plot of radiation energy density $u(x,\tau)$ as a function of space and time for $\epsilon =0$. Contrary to the results for finite $\epsilon$, the radiation energy density attains a finite value even at very early times due to the absence of retardation effects. However, the material energy density shows the same trend as for finite $\epsilon$. 

\subsection{Spherical shell}
\begin{figure}
\begin{center}
\includegraphics*[width=8cm]{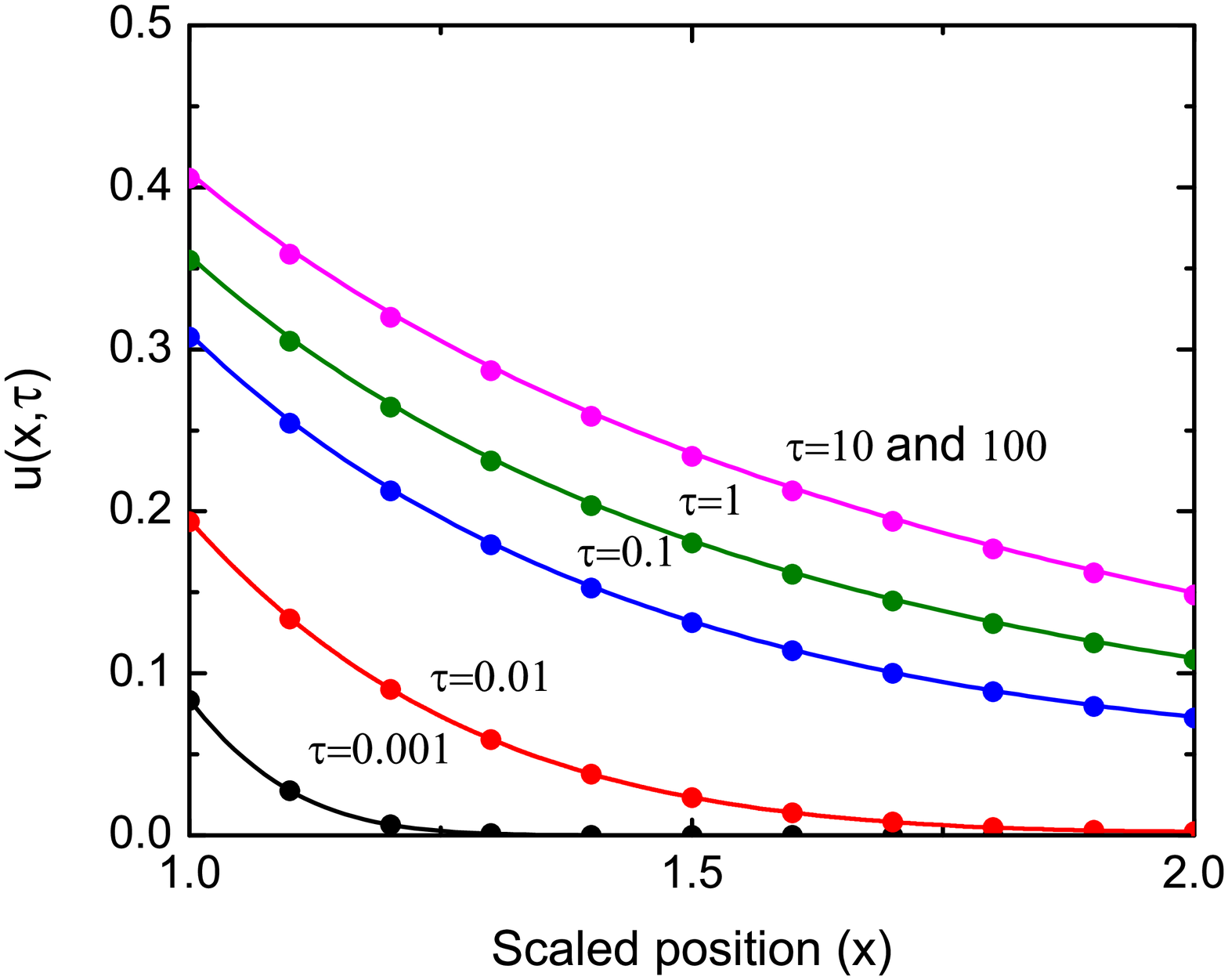}
\end{center}
\caption{\label{uxt-sp}Scaled radiation energy density $u(x,\tau)$ vs position (x) in a spherical shell of scaled inner radius $X_1=1$ and outer radius $X_2=2$ at different times for $ \epsilon = 0.1 $. The symbols stand for analytical values whereas lines represent the results obtained from finite difference method.}
\end{figure}

\begin{figure}
\begin{center}
\includegraphics*[width=8cm]{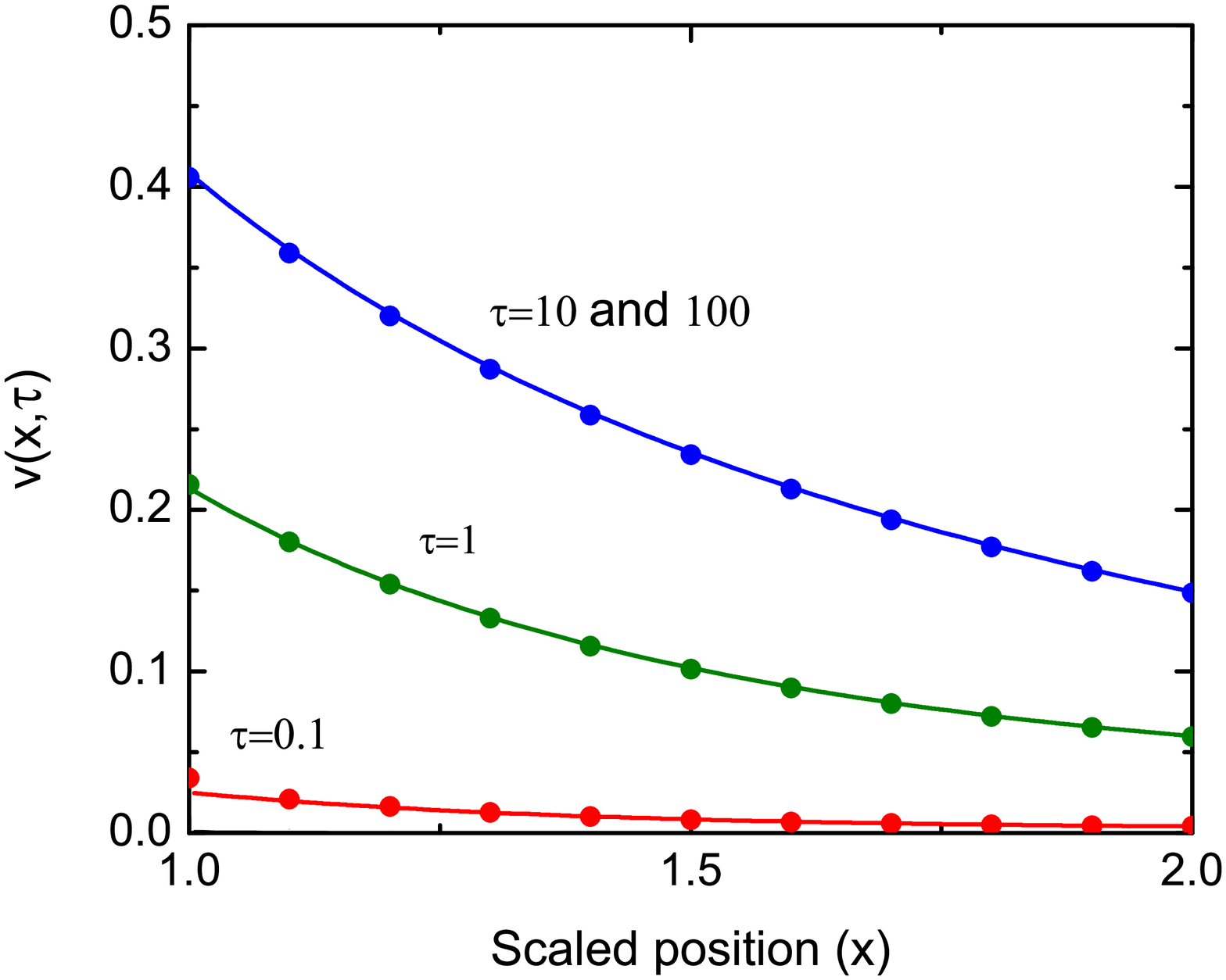}
\end{center}
\caption{\label{vxt-sp}Scaled material energy density $v(x,\tau)$ vs position in a spherical shell of scaled inner radius $X_1=1$ and outer radius $X_2=2$  at different times for $\epsilon = 0.1$. The symbols stand for analytical values whereas lines represent the results obtained from finite difference method.}
\end{figure}

\begin{figure}
\begin{center}
\includegraphics*[width=8cm]{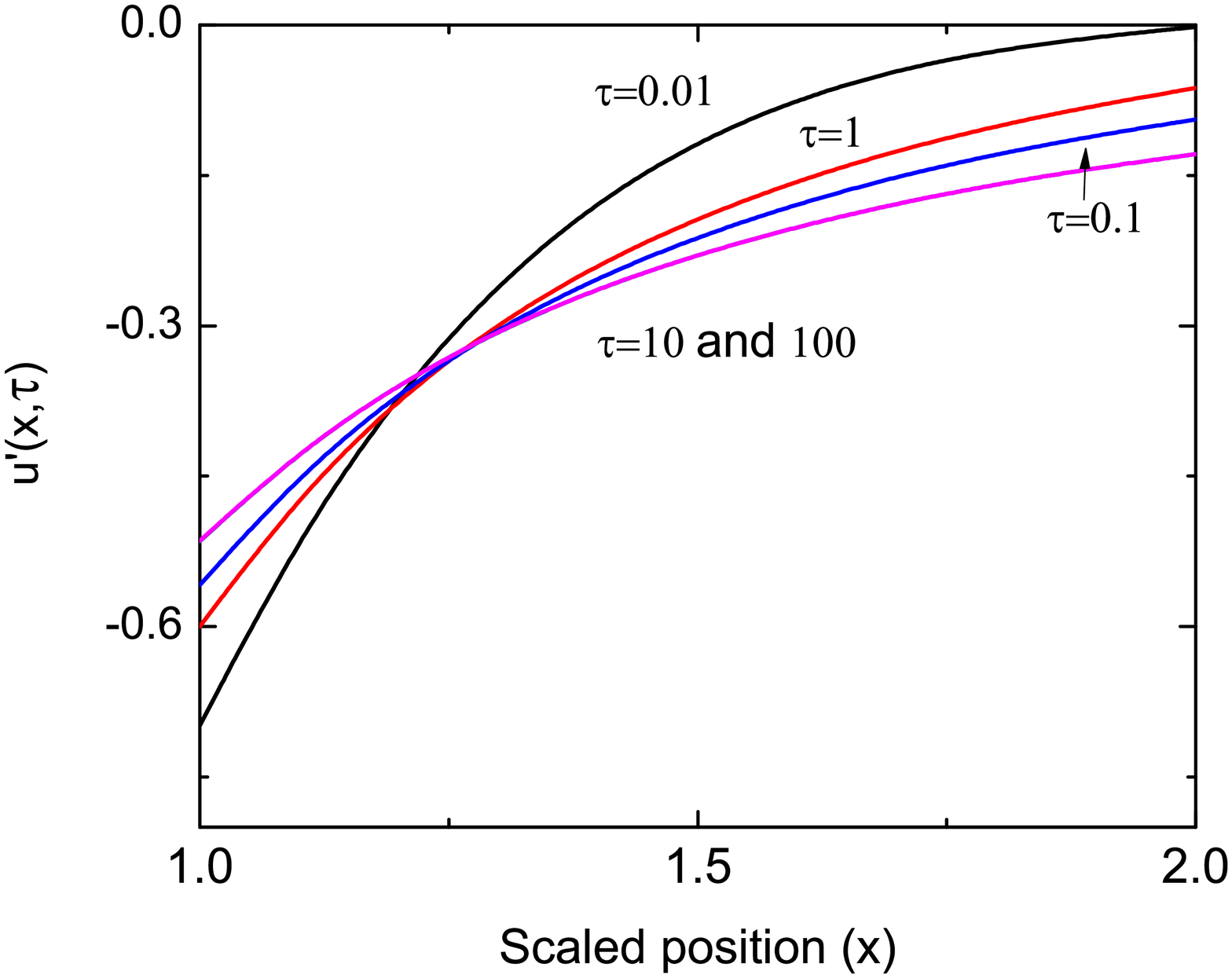}
\end{center}
\caption{\label{uprimext-sp}Space derivative of scaled radiation energy density $ u^\prime(x,\tau)$ vs position (x) in the spherical shell at different times.}
\end{figure}

\begin{figure}
\begin{center}
\includegraphics*[width=8cm]{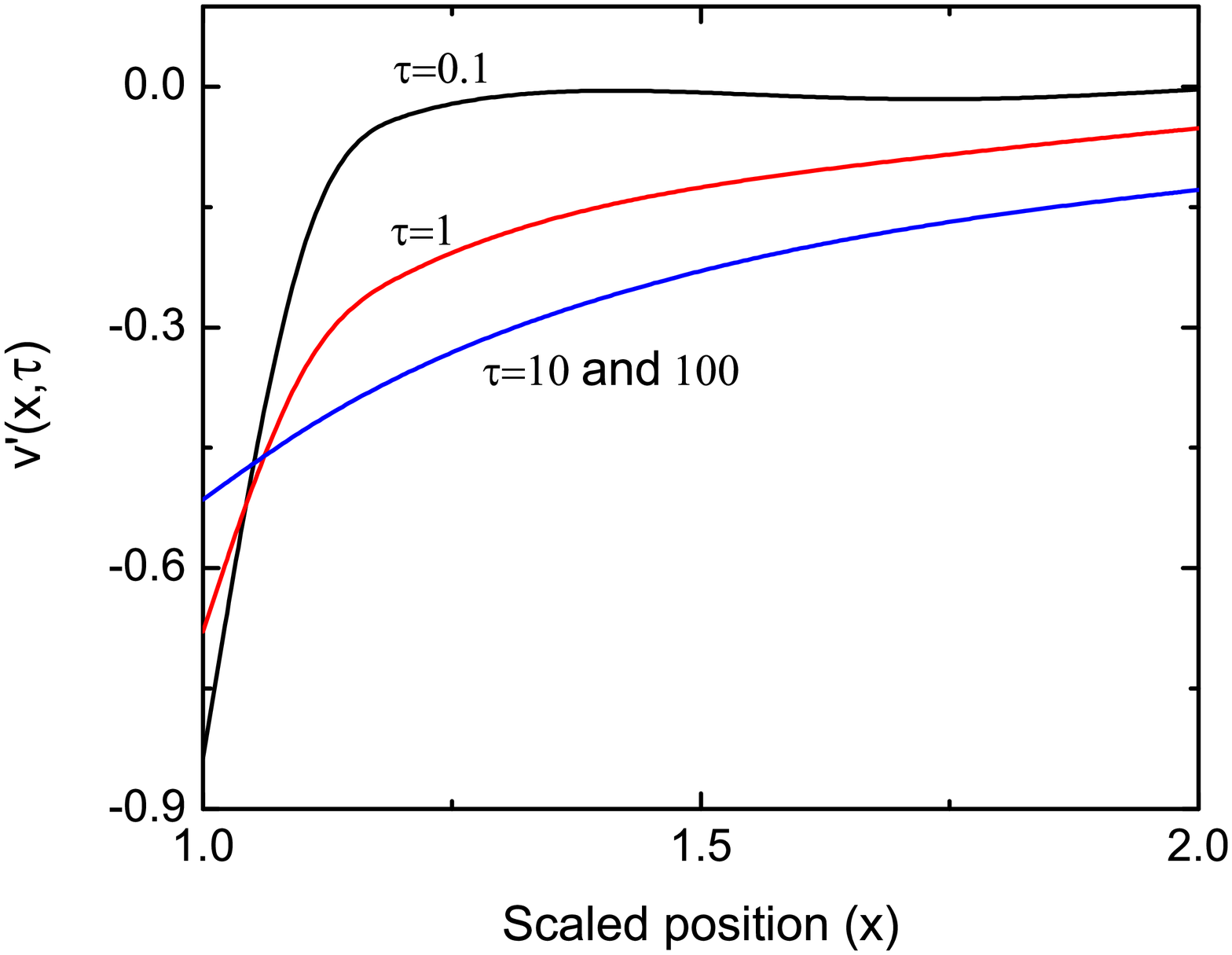}
\end{center}
\caption{\label{vprimext-sp}Space derivative of scaled material energy density $ v^\prime(x,\tau)$ vs position (x) in the spherical shell at different times.}
\end{figure}

\begin{figure}
\begin{center}
\includegraphics*[width=8cm]{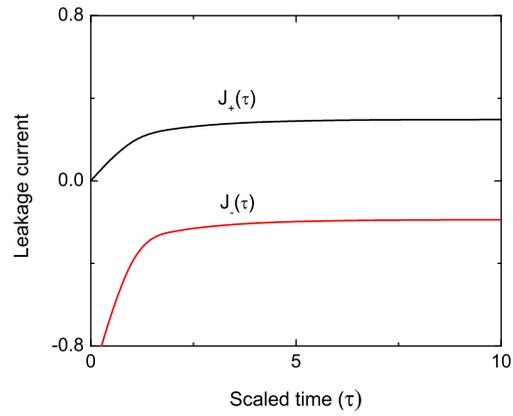}
\end{center}
\caption{\label{jplusminus-sp}Leakage currents $J_-(\tau)$ and $J_+(\tau)$ from the inner and outer surfaces of the spherical shell respectively.}
\end{figure}

\begin{figure}
\begin{center}
\includegraphics*[width=8cm]{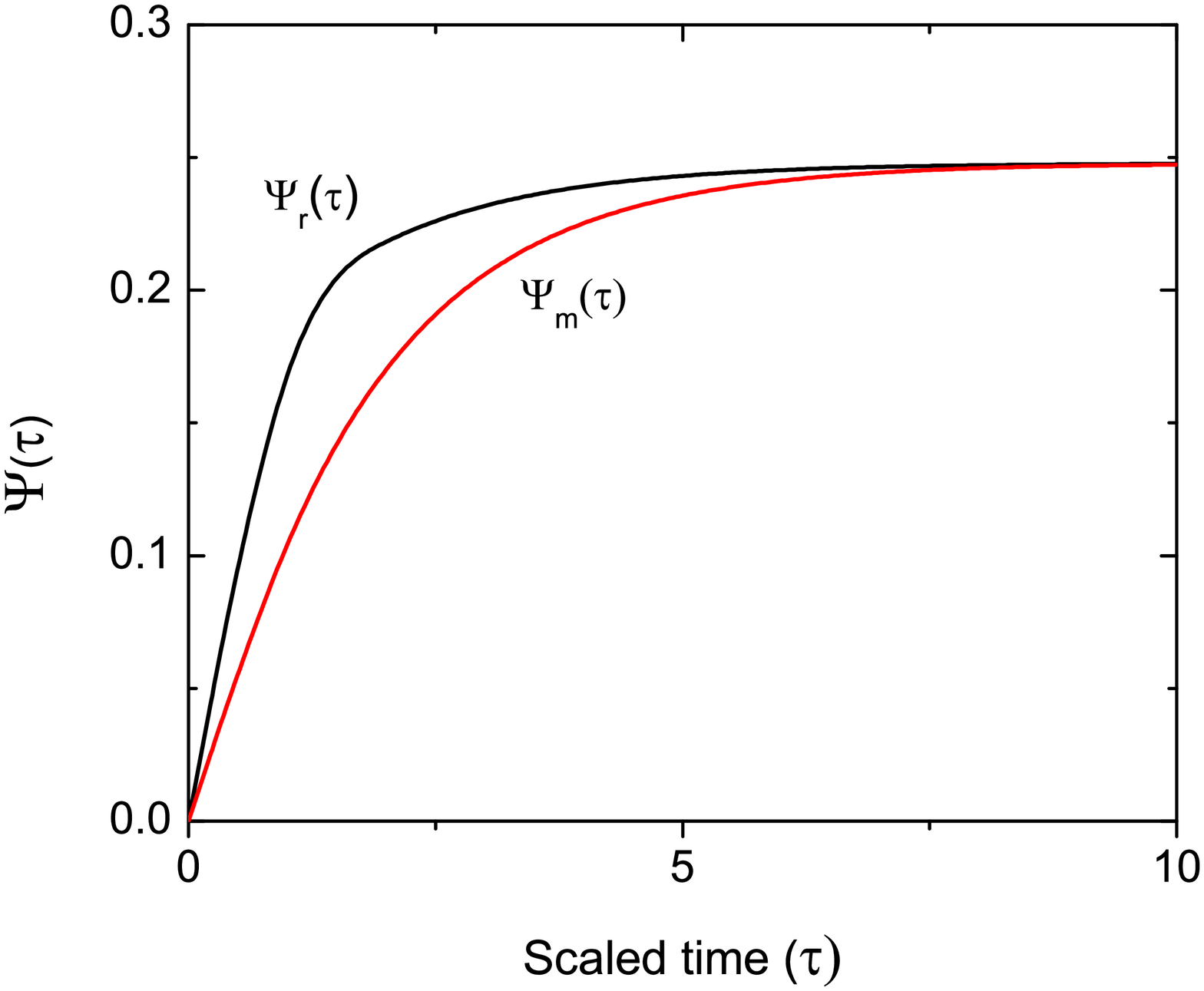}
\end{center}
\caption{\label{av-sp}Integrated radiation ($\psi_r(\tau)$) and material energy densities ($\psi_m(\tau)$) in the spherical shell as a function of scaled time $\tau$.}
\end{figure}

For the spherical shell, initially ($\tau$=0.01) the radiation energy density falls rapidly from the inner surface (scaled radius $X_1=1$) where radiation is incident towards the outer surface (scaled radius $X_2=2$) as shown in Fig. \ref{uxt-sp}. Though the trend is similar to the planar slab, the values of the scaled energy densities are less. Also, contrary to the planar case, the variation in energy densities remain sharper in the inner meshes compared to the outer ones and the variations in energy densities are not linear with position even after attaining steady state. This is evident because the mass of the material to be heated in the radially outward direction increases. Similar to the planar slab, the material energy density lags behind the radiation energy densities at early stages and finally reaches equilibrium (beyond $\tau$=10) [Fig. \ref{vxt-sp}]. Numerical results for energy densities in the spherical shell are obtained from finite difference analysis using the same mesh width and time step as used for the planar slab. Good agreement between the analytical and numerical results establish the validity of both the methods in spherical geometry. Magnitude of derivative of analytical radiation and material energy densities remains higher in the inner meshes as compared to outer ones at all times [Fig. \ref{uprimext-sp} and \ref{vprimext-sp}]. The leakage currents from the inner and outer surfaces of the spherical shell are $J_-(\tau) = u(X_1,\tau)+\frac{2}{\sqrt{3}}\frac{\partial u(X_1,\tau)}{\partial x},
J_+(\tau) = u(X_2,\tau)-\frac{2}{\sqrt{3}}\frac{\partial u(X_2,\tau)}{\partial x}$.
The variation in $J_+(\tau)$ is similar to planar slab though the values are less. However, $J_-(\tau)$ remains negative throughout as radiation always diffuses outwards in order to maintain the flux boundary conditions (Fig. \ref{jplusminus-sp}). As the derivative $\partial u(x, \tau)/\partial x$ is more negative for inner radii, $u(X_1, \tau)+u(X_2,\tau) <  1$ which leads to $J_+(\tau)+J_-(\tau) <  1$. For the case considered, $X_1=1$ and $X_2=2$, it is found that $2u(X_1,\tau) <  1$. As $J_-(\tau)=2u(X,\tau)-1$, hence $J_-(\tau)$ is negative.
The averaged or integrated radiation and material energy densities are given by $
\psi_r(\tau)=\int_{X_1}^{X_2} u(x,\tau)4\pi x^2dx$ and $\psi_m(\tau)=\int_{X_1}^{X_2} v(x,\tau)4\pi x^2dx$. 
and plotted in Fig. \ref{av-sp}. The integrated material energy density is also found to lag the radiation energy density at early times but finally the two equilibrate to a value of 0.25.
To check the consistency of the final results, we add Eqs. (\ref{RTE2s}) and (\ref{ME2s}) and integrate over x from $X_1$ to $X_2$, yielding 
\begin{eqnarray}
\int_{X_1}^{X_2}(\epsilon \frac{\partial u(x,\tau)}{\partial \tau}+\frac{\partial v(x,\tau)}{\partial \tau})4 \pi x^2 dx=4 \pi (X_2^2 \frac{\partial u(X_2,\tau)}{\partial x}-X_1^2 \frac{\partial u(X_1,\tau)}{\partial x}) 
\end{eqnarray}
Using the expressions for the energy densities, we find that both the left and right hand sides reduce to the same expression proving the consistency of the obtained solutions.

\begin{figure}
\begin{center}
\includegraphics*[width=8cm]{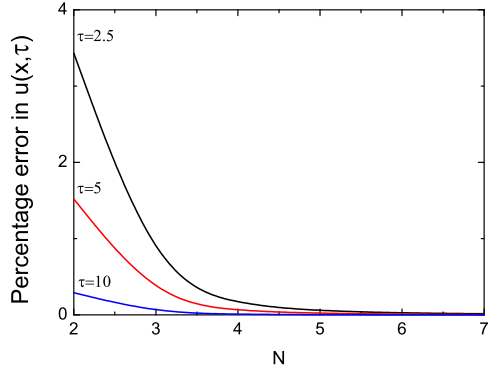}
\end{center}
\caption{\label{convergence-sph}Percentage error in the radiation energy density $u(x,\tau)$ in the spherical shell as a function of number of roots considered (N).}
\end{figure}

As for the planar slab, convergence of relative error in radiation energy density for spherical shell on increasing contribution from higher order poles is found to follow the same trend. However, the values of relative errors are slightly higher ($3.4\%$ for u(0,2.5) for contribution from first 2 roots) than the planar slab as shown in Fig. \ref{convergence-sph}. 
Thus for these finite systems, energy densities in terms of series solutions are found to converge quickly and depending on the required degree of accuracy, the number of poles to be considered is decided.

\section{Conclusions}\label{conc}
In this paper, the time dependent non equilibrium radiation diffusion problem has been solved analytically for finite planar slab and spherical shell with a constant radiation flux incident on the surface. The observed trend in temporal and spatial variation of energy densities, leakage currents, integral quantities, etc. has been explained physically. The analytical values of energy densities are cross checked with the solution of finite difference analysis and good agreement is observed when small mesh width and time steps are used. The results obtained in this paper can serve as new and useful benchmarks for non equilibrium radiation diffusion codes in both planar and spherical geometries. The same methodology can be applied to any other finite size systems like layered media with various boundary conditions. Using separation of variables, the method can be extended to generate benchmark results for validating radiation diffusion codes in two and three dimensions.
\\

{\bf Acknowledgement:}
\\
The author would like to thank Dr. N. K. Gupta for his useful comments, support and encouragement.


\end{document}